\documentclass[aps,prd,preprint,superscriptaddress,tightenlines,nofootinbib]{revtex4}

\usepackage{graphicx}% Include figure files
\usepackage{dcolumn}% Align table columns on decimal point
\usepackage{bm}% bold math

\begin{document}

%\preprint line(s) will be ignored for PRL/PRD
%\preprint{CLEO Draft YY-NNA} % For paper draft CBX YY-NN -> Draft YY-NNA
%\preprint{CLEO CONF YY-NN}   % For conference papers
%\preprint{ICHEP ABSnnn}      % For conference papers
\preprint{CLNS 02/1806}       % for CLNS notes
\preprint{CLEO 02-15}         % for CLNS notes

\title{First Search for the Flavor Changing Neutral Current Decay  
              $D^{0} \to \gamma\gamma$}
% for conference papers (ask CLEOAC for appropriate text)
%\thanks{Submitted to the 31$^{\rm st}$ International Conference on High Energy
%Physics, July 2002, Amsterdam}

%-------- INSERT HERE ------------
% Your author list goes here  REMOVE EVERYTHING to END INSERT and
% replace with your authorlist (ask cleoac).

\author{T.~E.~Coan}
\author{Y.~S.~Gao}
\author{F.~Liu}
\author{Y.~Maravin}
\author{R.~Stroynowski}
\affiliation{Southern Methodist University, Dallas, Texas 75275}
\author{M.~Artuso}
\author{C.~Boulahouache}
\author{S.~Blusk}
\author{K.~Bukin}
\author{E.~Dambasuren}
\author{G.~C.~Moneti}
\author{R.~Mountain}
\author{H.~Muramatsu}
\author{R.~Nandakumar}
\author{T.~Skwarnicki}
\author{S.~Stone}
\author{J.C.~Wang}
\affiliation{Syracuse University, Syracuse, New York 13244}
% ...
%\author{(CLEO Collaboration)} %FOR PRD_SPECIAL_CHANGEME
\collaboration{CLEO Collaboration} %FOR PRL,CLNS
\noaffiliation

%-------- END INSERT ------------

%please hard code the date when you have a final draft and submit to CLEOAC
\date{\today}

\begin{abstract} 

Using 13.8 fb$^{-1}$ of data collected at or just below the $\Upsilon(4S)$ 
with the CLEO  detector, we report the result of a search for the flavor 
changing neutral current process $D^{0}$ $\to$ $\gamma\gamma$.
We observe no significant signal for this decay mode and
determine 90\% confidence level upper limits on the branching fractions
${\cal B}(D^{0}\to \gamma\gamma)/{\cal B}(D^{0}\to\pi^{0}\pi^{0})$  $<$ 0.033 
and ${\cal B}(D^{0}\to\gamma\gamma )$  $<$ 2.9 $\times$ 10$^{-5}$.

\end{abstract}

\pacs{13.20.Fc}

\maketitle

In the standard model (SM), flavor changing neutral current (FCNC) processes
are forbidden at the tree level but can occur at higher loop level. Therefore,
they provide a good opportunity to probe new physics beyond the SM, especially
those processes where very small SM signals are expected.
The experimental studies of FCNC processes for charm have lagged behind 
those of the other flavors. 
The decay $D^{0}\to\gamma\gamma$ is a FCNC process which has not been 
measured. 
The branching fraction for $D^{0}\to\gamma\gamma$ expected from SM 
physics is about 10$^{-8}$ or less~\cite{fajfer,burdman},
but gluino exchange in supersymmetric models with reasonable 
parameters can enhance the SM rate by as much as two orders
of magnitude \cite{prelovsek-wyler}.

In this Letter, we present results of the first search for the FCNC process  
$D^{0}\to\gamma\gamma$. 
The data were collected with two configurations 
(CLEO II~\cite{CLEOII} and CLEO II.V~\cite{CLEOII.V}) of the
CLEO detector at the Cornell Electron Storage Ring (CESR). 
They consist of 13.8 fb$^{-1}$ taken at or just below the $\Upsilon$(4S)
where $c\bar{c}$ or other accessible quark pairs are produced with the 
$D^{0}$ candidates produced in the hadronization of $c\bar{c}$ pairs.
The final states of the decays under study are reconstructed by
combining detected photons or neutral pions with charged pions. 
The detector elements most important for the results presented 
here are the tracking system, which consists of several  
concentric detectors operating inside a 1.5 T superconducting solenoid, 
and the high-resolution electromagnetic calorimeter, consisting of 7800 CsI(Tl)
crystals. For CLEO II, the tracking system consists of a 6-layer
straw tube chamber, a 10-layer precision drift chamber, and a
51-layer main drift chamber. The main drift chamber also provides a
measurement of the specific ionization loss, $dE/dx$, used for
particle identification.  For CLEO II.V the straw tube
chamber was replaced by a 3-layer, double-sided silicon vertex
detector, and the gas in the main drift chamber was changed from 
an argon-ethane to a helium-propane mixture.

Due to the small expected branching fraction, the mass resolution
for two-photon final state, and the huge combinatoric backgrounds from random 
photons, it is extremely difficult to find a $D^{0}$ mass peak in 
$\gamma\gamma$ invariant mass if searching directly for 
$D^{0}\to\gamma\gamma$. 
However, the situation is very different if we require the $D^{0}$ to
be produced from the decay $D^{*+} \to D^{0}\pi^{+}$.
The energy release $Q$ of candidates for the decay
$D^{*+} \to D^{0} \pi^+$ is given by
$Q/c^2 \equiv M(D^{*+}) - M(D^0) - M_{\pi^+}$.
In this expression, $M(D^{*+})$ and $M(D^0)$ are the invariant masses of 
the $D^{*+}$ and $D^0$ candidates, respectively, and $M_{\pi^+}$ is 
the $\pi^+$ mass \cite{pdg}.
There exists an excellent calibration mode, 
$D^{*+} \to D^{0}\pi^{+}$ where $D^{0} \to \pi^{0}\pi^{0}$.
This calibration mode and the signal mode have similar final state 
particles so many common systematic errors will cancel.
The branching fraction of $D^{0} \to \pi^{0}\pi^{0}$ was measured to be
${\cal B}( D^{0} \to \pi^{0}\pi^{0} )$ $=$ (8.4 $\pm$ 2.2) $\times$ 
10$^{-4}$~\cite{pdg,selen}.
The CLEO resolution in $Q$ is better than 1 MeV and does not differ
significantly between the two $D^0$ decay modes. 
The analysis strategy to search for $D^{0} \to \gamma\gamma$ is to use 
$D^{*+} \to D^{0}\pi^{+}$ where $D^{0} \to \gamma\gamma$ and 
$D^{0} \to \pi^{0}\pi^{0}$, and calculate the ratio,
${\cal B}(D^{0}\to \gamma\gamma)/{\cal B}(D^{0}\to\pi^{0}\pi^{0})$.
Charge-conjugate modes are implied throughout this Letter.

Candidates for $D^{0}$ meson decays are reconstructed by combining two 
detected photons or neutral pions. 
The invariant mass of the two photons or neutral pions is required to be 
within 2.5 standard deviations ($\sigma$) of the known 
$D^{0}$ mass~\cite{pdg}. 
The photon candidates are required to pass quality cuts and not to be 
associated with charged tracks.
To form $\pi^0$ candidates, pairs of photon candidates with invariant mass 
within $3\sigma$ of the $\pi^0$ mass $M_{\pi^{0}}$~\cite{pdg} are fitted
kinematically with the mass constrained to $M_{\pi^{0}}$.
To reduce combinatoric backgrounds, each $\pi^{0}$ or photon candidate in
the $D^{0} \to \pi^{0}\pi^{0}$ or $\gamma\gamma$ candidate is required to have
momentum greater than 0.55 GeV/$c$. Furthermore, each $D^{0}$ candidate is
required to have momentum greater than 2.2 GeV/$c$. These requirements come
from an optimization that minimizes the statistical error of the branching 
fraction $D^{*+} \to D^{0}\pi^{+}$ where $D^{0} \to \pi^{0}\pi^{0}$.

A $\pi^{+}$ is then combined with the $D^{0}$ candidate to form a
$D^{*+}$ candidate. The $\pi^{+}$ candidate must be a well-reconstructed track
originating from a cylinder of radius 3 mm and half-length 5 cm centered
at the $e^{+}e^{-}$ interaction point. The minimum momentum requirement on
the $D^{0}$ candidate of 2.2 GeV/$c$ corresponds to a lower limit on the 
$\pi^{+}$ momentum of approximately 100 MeV/$c$.
The $dE/dx$ information for the $\pi^+$ candidate, 
if it exists and is reliable, 
is required to be within $3\sigma$ of its expected value.
The two energetic photons, one from each $\pi^{0}$ in 
$D^{0}\to\pi^{0}\pi^{0}$ decay, can form fake $D^{0} \to \gamma\gamma$ 
candidate. 
To limit cross-feed from $D^{0} \to \pi^{0}\pi^{0}$, a photon 
candidate in $D^{0} \to \gamma\gamma$ is rejected if $M(\gamma\gamma )$
is within $3\sigma$ of $M_{\pi^{0}}$ when combined with any other photon
in the event ($\pi^{0}$ veto).

To estimate the detection efficiencies and backgrounds, we generate
$D^{*+}\to D^{0}\pi^{+}$ with $D^{0} \to \gamma\gamma$ and 
$D^{0} \to \pi^{0}\pi^{0}$, together with generic Monte Carlo events
($u\bar{u}$, $d\bar{d}$, $s\bar{s}$, and $c\bar{c}$) 
produced near the $\Upsilon(4S)$, 
and  simulate the CLEO detector response with GEANT~\cite{GEANT}.  
Simulated events for the CLEO II and CLEO II.V 
configurations are processed in the same manner as the data. 
With the above event selection, multiple candidates per event are rare
(less than 1\%).
From Monte Carlo simulations, the cross feed contribution from 
$D^{0} \to \pi^{0}\pi^{0}$ to $D^{0} \to \gamma\gamma$ in the signal region 
of the $Q$ distribution is about one event (or about 4 events without
the $\pi^{0}$ veto). 
Other cross feed contributions from possible $D^{0} \to \eta\eta$,
$\eta X$ decays are negligible.

Figure 1 shows the $Q$ distributions for
$D^{*+} \to D^{0}\pi^{+}$ candidates where $D^{0} \to \pi^{0}\pi^{0}$ and
$D^{0} \to \gamma\gamma$. The circles with error bars are CLEO data
which are fit using a binned likelihood fit to a Gaussian function with 
expected mean and width determined from signal Monte Carlo simulation, 
on top of a threshold background function.
For $D^{*+}\to D^{0}\pi^{+}$ where $D^{0} \to \pi^{0}\pi^{0}$, 
$N(\pi^0\pi^0) = 628.0 \pm 31.8$ signal events are observed. 
The signal and background levels found in the data are in good agreement 
with those obtained from Monte Carlo simulations.
For $D^{*+}\to D^{0}\pi^{+}$ where $D^{0} \to \gamma\gamma$, no significant
enhancement is observed in the signal region. The signal yield from the fit 
is $N(\gamma\gamma) = 19.2 \pm 9.3$ events.

\begin{figure}
\includegraphics*[width=3.50in]{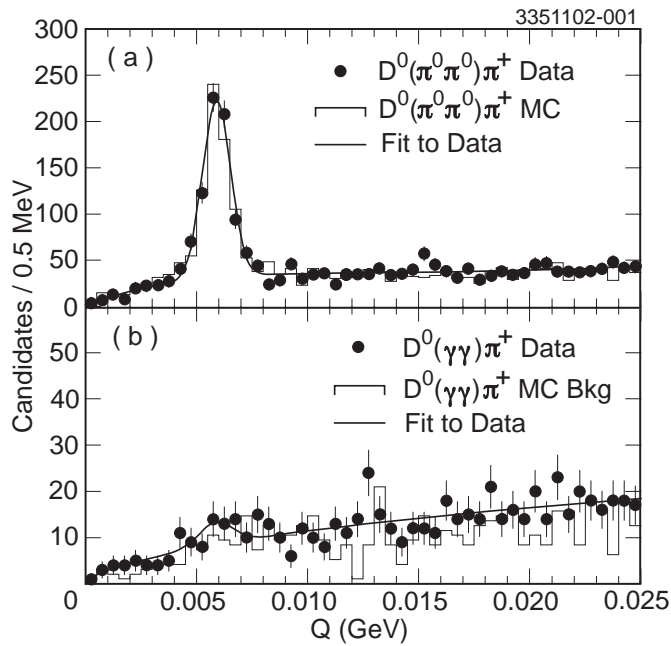}
\caption{The $Q$ distributions for $D^{*+} \to D^{0}\pi^{+}$ where 
         $D^{0} \to \pi^{0}\pi^{0}$ (a) and $D^{0} \to \gamma\gamma$ (b). 
         The circles are from data and the solid curves are the fit to the 
         data.
         The histograms are from the normalized Monte Carlo expectations.}
\label{fig01}
\end{figure}

From Monte Carlo simulations, 
the relative efficiency for $D^{0} \to \gamma\gamma$ and
$D^{0} \to \pi^{0}\pi^{0}$ is determined to be:
$\epsilon(\gamma\gamma )/\epsilon(\pi^{0}\pi^{0})$ = 1.58 $\pm$ 0.05.
The systematic uncertainties come from event selection, signal yield from
data and Monte Carlo simulation, and are listed in Table I.
The other common systematic uncertainties for $D^{*+} \to D^{0}\pi^{+}$ cancel 
in measuring 
${\cal B}(D^0 \to \gamma\gamma)/{\cal B}(D^0 \to \pi^0\pi^0) =
      [N(\gamma\gamma)/N(\pi^0\pi^0)]/
      [\epsilon(\gamma\gamma)/\epsilon (\pi^0\pi^0)] = 0.0194 \pm 0.0094 $.
Combining the signal yields, relative selection efficiency and systematic
uncertainties in $D^{*+} \to D^{0}\pi^{+}$ where $D^{0} \to \pi^{0}\pi^{0}$ and
$D^{0} \to \gamma\gamma$, we then obtain a 90\% confidence level (C.L.)
upper limit by the following method.
We only consider the physical region of the ratio 
${\cal B}(D^{0}\to\gamma\gamma )/{\cal B}(D^{0}\to\pi^{0}\pi^{0})$ 
assuming that the shape of the likelihood is Gaussian with an unknown mean, 
but whose standard deviation is determined by the statistical and systematic 
errors added in quadrature. We then determine the 90\% C.L. upper limit to be 
the mean of the Gaussian, 90\% of whose probability lies above the
observed ratio.
We set an upper limit:
${\cal B}(D^{0}\to\gamma\gamma )/{\cal B}(D^{0}\to\pi^{0}\pi^{0})$ 
$<$ 0.033 at the 90\% C.L.
Using ${\cal B}(D^{0}\to\pi^{0}\pi^{0} )$ $=$ (8.4 $\pm$ 2.2) $\times$ 
10$^{-4}$~\cite{pdg,selen}, we similarly set an upper limit:
${\cal B}(D^{0}\to\gamma\gamma )$ $<$ 2.9 $\times$ 10$^{-5}$ at the 90\% C.L.

In summary, we report the result of a search for the FCNC process 
of $D^{0} \to \gamma\gamma$.
We observe no significant signal for this decay mode and determine the
first 90\% C.L. upper limits on the following branching fractions:
${\cal B}(D^{0}\to \gamma\gamma)/{\cal B}(D^{0}\to\pi^{0}\pi^{0})$  $<$ 0.033 
and ${\cal B}(D^{0}\to\gamma\gamma )$  $<$ 2.9 $\times$ 10$^{-5}$.
This result is an order of magnitude above the theoretical prediction
of Ref.~\cite{prelovsek-wyler}.

We gratefully acknowledge the effort of the CESR staff in providing us 
with excellent luminosity and running conditions.
We appreciate useful discussions with M. Neubert.
This work was supported by the National Science Foundation,
the U.S. Department of Energy, the Research Corporation,
and the Texas Advanced Research Program.

\begin{table}[hhh]
\label{systmatic_table}
\caption{Summary of systematic error sources and their contribution in
         measuring the ratio
         ${\cal B}(D^{0}\to\gamma\gamma )/{\cal B}(D^{0}\to\pi^{0}\pi^{0})$.}
 \begin{center}
 \begin{tabular}{c c} \hline \hline
  Systematic Error Source        & Error (\%)                   \\  \hline 
  $\pi^{0}$ finding efficiency 
                                 &    $5.0/\pi^0$                \\  
  $\gamma$ finding efficiency     
                                 &    $3.0/\gamma$               \\ 
  Fit Yield                    
                                 &    3.0                        \\  
  $D^{0}$ selection             
                                 &    2.0                        \\  
  MC Statistics 
                                 &    2.0                        \\   
  Hadronic Event Selection       
                                 &    1.0                        \\  \hline
  Total for $D^{0}\to\gamma\gamma$       
                                 &    7.3                        \\  \hline
  Total for $D^{0}\to\pi^{0}\pi^{0}$       
                                 &    10.9                       \\  \hline
  Total for       
    ${\cal B}(D^{0}\to\gamma\gamma )/{\cal B}(D^{0}\to\pi^{0}\pi^{0})$ 
                                 &    13.1            \\  \hline  \hline
\end{tabular}
\end{center}
\end{table}

\end{document}